\documentclass[aps,prx,twocolumn,superscriptaddress,nofootinbib]{revtex4-1}
\usepackage[T1]{fontenc}
\usepackage{graphicx}
\usepackage{epstopdf}
\usepackage{gensymb}
\usepackage{amsmath}
\usepackage{units}
\usepackage[colorlinks=true, allcolors={black}]{hyperref}
\usepackage{upgreek}
\usepackage{scrextend}

\usepackage[utf8]{inputenc}

\begin{document}

\preprint{}

\title[\hspace{20em}]{Waveguide-integrated van der Waals heterostructure photodetector at telecom band with high speed and high responsivity}




\author{Nikolaus Fl\"ory}
\thanks{These authors contributed equally.}
\affiliation{Photonics Laboratory, ETH Z{\"u}rich, 8093 Z{\"u}rich, Switzerland}

\author{Ping Ma}
\thanks{These authors contributed equally.}
\affiliation{Institute of Electromagnetic Fields, ETH Z{\"u}rich, 8092 Z{\"u}rich, Switzerland}

\author{Yannick Salamin}
\affiliation{Institute of Electromagnetic Fields, ETH Z{\"u}rich, 8092 Z{\"u}rich, Switzerland}

\author{Alexandros Emboras}
\affiliation{Institute of Electromagnetic Fields, ETH Z{\"u}rich, 8092 Z{\"u}rich, Switzerland}

\author{Takashi  Taniguchi}
\affiliation{National Institute for Material Science, 1-1 Namiki, Tsukuba 305-0044, Japan}

\author{Kenji Watanabe}
\affiliation{National Institute for Material Science, 1-1 Namiki, Tsukuba 305-0044, Japan}

\author{Juerg Leuthold}
\email{leuthold@ethz.ch}
\affiliation{Institute of Electromagnetic Fields, ETH Z{\"u}rich, 8092 Z{\"u}rich, Switzerland}

\author{Lukas Novotny}
\email{lnovotny@ethz.ch}
\affiliation{Photonics Laboratory, ETH Z{\"u}rich, 8093 Z{\"u}rich, Switzerland}

\email{$^{\ast}$ mapi@ethz.ch, leuthold@ethz.ch, lnovotny@ethz.ch} 

\date{\today}

\begin{abstract}

\end{abstract}

\maketitle 

\textbf{Intensive efforts have been devoted to exploit novel optoelectronic devices based on two-dimensional (2D) transition-metal dichalcogenides (TMDCs) owing to their strong light-matter interaction and distinctive material properties~\cite{Wang2012,Xiao2017}. In particular, photodetectors featuring both high-speed and high-responsivity performance are of great interest for a vast number of applications such as high-data-rate interconnects operated at standardized telecom wavelengths ~\cite{Koppens2014a,Ferrari2018}. Yet, the intrinsically small carrier mobilities of TMDCs become a bottleneck for high-speed application use ~\cite{Konstantatos2018}. Here, we present high-performance vertical van der Waals heterostructure-based photodetectors integrated on a silicon photonics platform. Our vertical MoTe$_2$/graphene heterostructure design minimizes the carrier transit path length in TMDCs and enables a record-high measured bandwidth of at least 24\,GHz under a moderate bias voltage of $-$3 volts. Applying a higher bias or employing thinner MoTe$_2$ flakes boosts the bandwidth even to 50\,GHz. Simultaneously, our device reaches a high external responsivity of 0.2\,A/W for incident light at 1300\,nm, benefiting from the integrated waveguide design. Our studies shed light on performance trade-offs and present design guidelines for fast and efficient devices. The combination of 2D heterostructures and integrated guided-wave nano photonics defines an attractive platform to realize high-performance optoelectronic devices~\cite{Mueller2015,Liu2016f,Yu2013e}, such as photodetectors~\cite{Britnell2013b}, light-emitting devices~\cite{Withers2015} and electro-optic modulators~\cite{Sun2016b}. }\\

During the last decade, two-dimensional (2D) materials such as graphene and transition-metal
dichalcogenides (TMDCs) have shown great promise for a wide range of photonic and optoelectronic applications~\cite{Bonaccorso2010a,Boltasseva2019,Manzeli2017}. 2D devices have the potential to outperform established and more mature technologies, particularly in terms of form factor, operating conditions and cost-effectiveness. The possibility to integrate 2D materials without constraints of crystal lattice matching is disruptive, as it tremendously simplifies manufacturing and increases possible material combinations. 
Graphene, which has been widely used for successful 2D device implementations~\cite{Kim2011,Schuler2018,Phare2015,Gan2013d,Hone2015,Ma2019}, has an intrinsically weak photosensitivity, though its interaction with light can be enhanced using silicon-based integrated photonics, such as optical resonators~\cite{Phare2015} or waveguides~\cite{Youngblood2016a}. Nevertheless, graphene-based devices suffer from other issues stemming from its gapless nature, e.g., large dark currents for photodetectors. Alternatively, TMDCs, a semiconducting class of 2D materials, hold great promise for high-performance optoelectronic devices due to their intrinsically strong light-matter interactions~\cite{Mak2016a}. Yet, the integration with a silicon-based platform is challenging, because direct band-to-band transition energies of TMDCs fall within the absorption band of silicon. Despite of this, few attempts have been made towards the integration of TMDCs with silicon-based structures~\cite{Bie2017,Ma2018a}, but high-performance devices, especially operated at standardized telecom bands relevant for applications of information and communication technology (ICT)~\cite{Ferrari2018}, are still under very limited investigation.

Another major challenge of employing TMDCs for ICT devices is the speed performance. This is of particular importance when TMDCs are employed in photodetectors ~\cite{Koppens2014a,Konstantatos2018,Buscema2015}. While impressive photoresponsivities and photoconductive gains have been demonstrated~\cite{Lopez-Sanchez2013a,Yu2017}, what is easily overseen is that these high gains typically originate from photogating effects~\cite{Konstantatos2018,Lopez-Sanchez2013a}. The associated long carrier lifetimes inherently limit the speed performance~\cite{Konstantatos2018,Wang2016h}. Moreover, the carrier mobilities of TMDCs are significantly smaller than those of graphene~\cite{Ma2018a,Octon2016c}, which poses an obstacle for high-speed device performance. To date, the highest reported bandwidth of TMDC waveguide detectors is below 1\,GHz~\cite{Bie2017,Ma2018a}, limited by the relatively long transit time of carriers before they are collected. It is therefore highly desirable to investigate TMDC device configurations for an improved speed performance, e.g. comparable to graphene-based devices, but with higher efficiency.

In this Letter we present high-speed and high-responsivity vertical MoTe$_2$/graphene van der Waals heterostructure photodetectors integrated with planar silicon photonic waveguides to address the aforementioned key technology challenges. The device design takes full advantage of van der Waals heterostructures as well as of the waveguide integration scheme. Both of these are essential to overcome the intrinsic material constraints and allow us to demonstrate application-level device performance of up to 50\,GHz bandwidth and photoresponsivities of more than 0.2\,A/W with low dark currents at the telecom band.

\begin{figure*}[t!]
\begin{center}
\includegraphics[width=\textwidth]{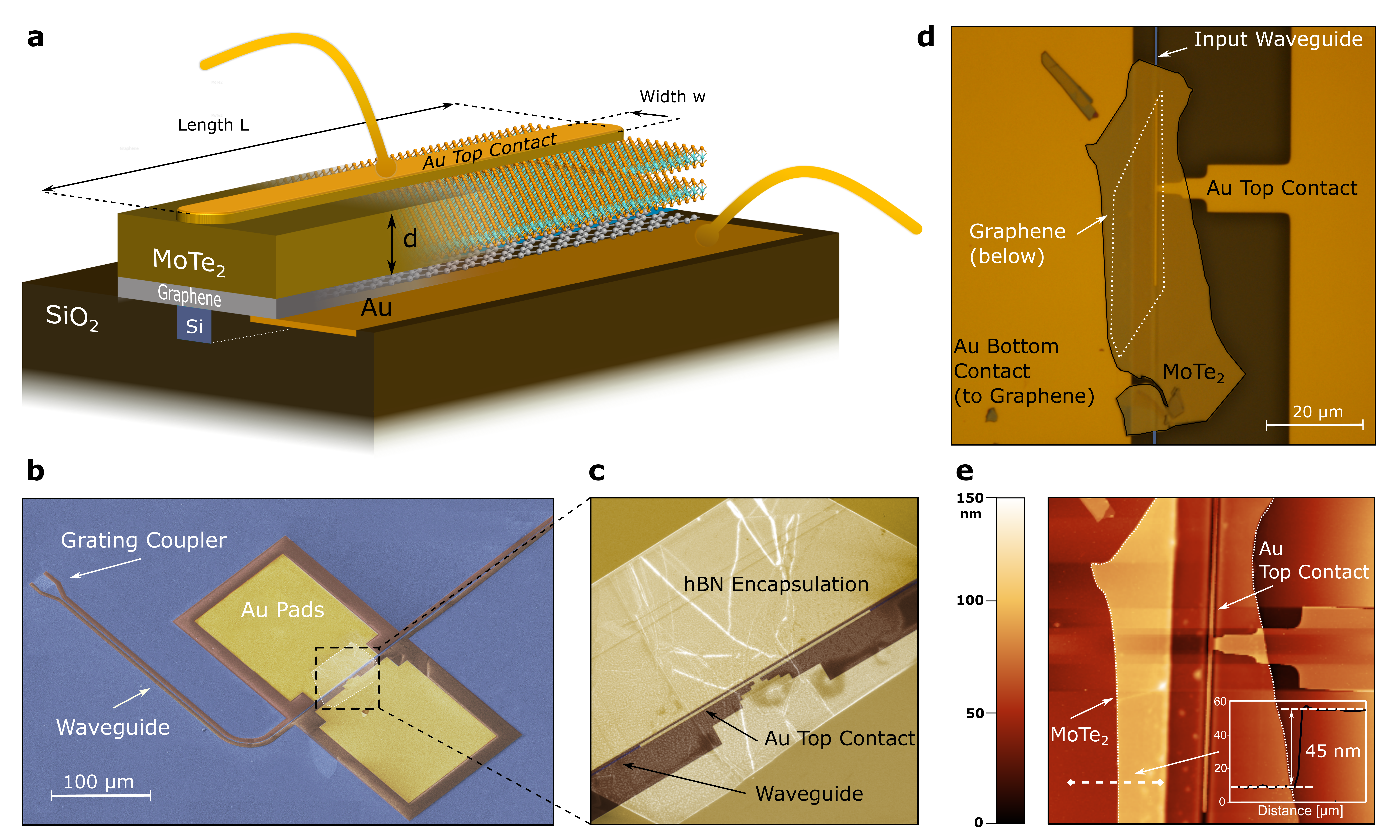}
\caption{\textbf{Vertical MoTe$_2$-graphene photodetector.} \textbf{a}, Schematic illustration of a vertical
MoTe$_2$-graphene heterostructure detector coupled to a silicon waveguide buried in SiO$_2$
claddings. Graphene and MoTe$_2$ are connected to gold (Au) bottom and top
contacts, respectively. \textbf{b}, False-color scanning electron microscope (SEM) image of a
fabricated device, showing the silicon waveguide and grating coupler (GC, both in blue color),
the waveguide oxide lateral claddings (in brown color), the metallic structures (in golden
color), and the encapsulation hBN layer (in semi-transparent white color). \textbf{c}, Enlarged-view SEM image of the fabricated detector. \textbf{d}, Optical micrograph picture of a
fabricated waveguide detector prior to encapsulation with hBN. It shows the graphene flake, the
MoTe$_2$ flake, the optical waveguide and metallic
structures, including the contact electrode on top of the MoTe$_2$, used for carrier extraction. \textbf{e}, Atomic force
microscope (AFM) image of the fabricated detector, showing the MoTe$_2$ flake and the metallic
contact bar on top of the MoTe$_2$ flake and the planar waveguide. Inset: a cross-sectional line-scan indicating a MoTe$_2$ thickness of 45\,nm.
\label{fig1}}
\end{center}
\end{figure*}

\begin{figure*}[t!]
\begin{center}
\includegraphics[width=\textwidth]{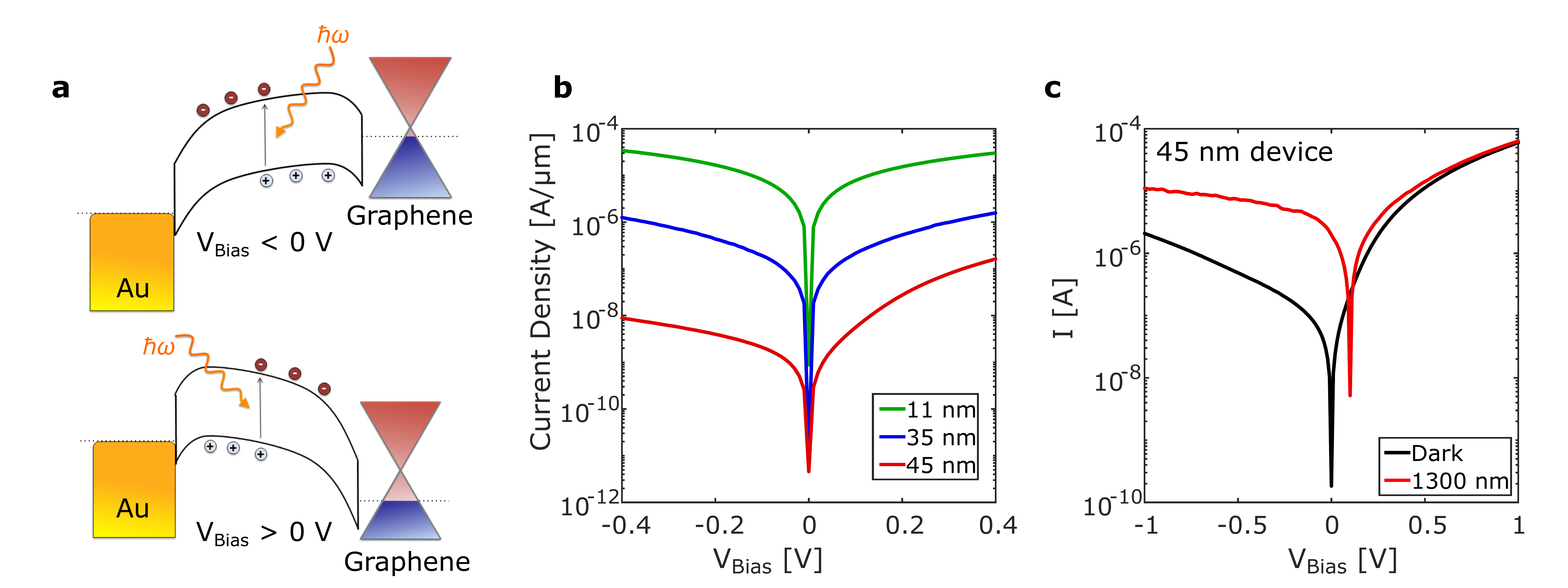}
\caption{\textbf{Electrical Characteristics of the photodetector.} \textbf{a}, Schematic illustration of the band diagrams of the vertical MoTe$_2$-graphene heterostructure under negative bias (upper diagram) and positive bias
(lower diagram), respectively. \textbf{b}, Dependence of the current density on bias voltage on a logarithmic scale for several devices with different MoTe$_2$ channel thicknesses (red line, 45\,nm; blue line, 35\,nm; green line, 11\,nm). The current density is obtained by normalizing the current to the length of
each device. \textbf{c}, Current-voltage (I-V) curves without (black curve) and with (red curve) 1300\,nm light coupled to a detector featuring a 45\,nm thick MoTe$_2$ flake.
\label{fig2}}
\end{center}
\end{figure*}

Figure~\ref{fig1}a illustrates the concept of the presented photodetector design. A thin flake of
semiconducting few-layer MoTe$_2$ is introduced as the light absorbing medium. MoTe$_2$ is compatible with silicon integrated photonics, since it exhibits a layer-dependent bandgap and strong light absorption extending into the
standard telecom O-band wavelength range (1260-1360\,nm)~\cite{Ma2018a,Ruppert2014}. The MoTe$_2$ flake is vertically sandwiched between two parallel electrodes in
order to build up a vertical carrier drift path of short distance. The device is operated with transverse electric
(TE) polarized light, which has its dominant electric field component parallel to the plane
of the MoTe$_2$ flake so as to be efficiently absorbed via band-to-band transitions. Light propagating in the silicon waveguide overlaps evanescently with the absorbing MoTe$_2$ in the active section of the detector. Electron-hole pairs generated by the
absorbed photons are efficiently separated and extracted by the uniform electric field applied between the bottom and top electrodes. 

The high speed of the device is a result of several factors. First, the use of a vertical heterostructure remedies the low mobility of TMDCs. The bandwidth ($f_{3dB}$) of a transit-time-limited
device is $f_{3dB} \sim 0.55/\tau_{tr} = 0.55 \cdot \nu /L$ where $\tau_{tr}$ is the carrier transit time, $\nu$ the carrier velocity and $L$ the length of the carrier transit path~\cite{Kato1993,Xia2009b,Massicotte2015c}. In previously reported TMDC waveguide detectors, the photo-excited carriers were transported in-plane over much longer distances by lateral electric fields~\cite{Bie2017,Ma2018a,Youngblood2015}.
A vertical heterostructure on the other hand enables a
vertical channel that restricts the transit path length of photoexcited carriers down to a few nanometers, thereby achieving much smaller transit times. Second, monolayer graphene is adopted as a transparent bottom electrode. On one hand, it allows the optical mode to spatially overlap with the
MoTe$_2$ absorber and, on the other hand, the high conductivity and carrier mobility
of graphene ensures fast carrier extraction and small series resistance. Third, a narrow metallic contact made of thin-film gold (Au) on top of MoTe$_2$, is aligned along the integrated waveguide and results in a small but well-defined vertical carrier extraction channel overlapping with the guided optical mode. The use of such a shaped top electrode allows the active area of the device to be very small, which minimizes the circuital capacitance of the device. Together with the small circuital resistance, this results in a large resistance-capacitance (RC)-limited bandwidth. Besides, the asymmetric contact scheme generates a built-in electric field that contributes to the carrier separation under zero and low bias conditions.

It is noteworthy that by design the light absorption and detection of the presented devices
are scalable with the length of the integrated waveguide, while the collection path of
photogenerated carriers is perpendicular to the light propagation direction. As a result, our
device has no trade-offs between the carrier transit-time limited bandwidth and quantum
efficiency. This is different for surface-normal illuminated photodetectors, for which the length of the
absorber is finite and light absorption is low. The waveguide integration
approach in our device is crucial for achieving high-speed performance while maintaining a high
responsivity.

The fabrication of our devices is detailed in the Methods and Supplementary Section S1. In short, buried silicon waveguides were first fabricated on standard silicon-on-insulator wafers by using a LOCal Oxidation of Silicon (LOCOS) technique~\cite{Desiatov2010,Naiman2015a}. Grating couplers (GCs) were produced by shallow etching of silicon. Flakes of MoTe$_2$ with various thicknesses and monolayer graphene were obtained by means of mechanical exfoliation and stacked employing a polymer-based pick-up technique~\cite{Britnell2013b,Zomer2014a}. The stacked MoTe$_2$-graphene heterostructures were subsequently transferred to the silicon photonics chips and aligned to the buried waveguides.
After transfer of the 2D flakes, a top metallic contact was patterned onto the MoTe$_2$ by
electron-beam lithography (EBL), metal evaporation and a lift-off process. Figure ~\ref{fig1}b displays
a scanning electron microscope (SEM) image of a fabricated device, including the access
silicon waveguide, the GC, and the detector structure as shown in Fig.~\ref{fig1}c with an enlarged
view. Figure~\ref{fig1}d and ~\ref{fig1}e show a top-view optical microscope image and an atomic force
microscope (AFM) image of a fabricated detector before final device encapsulation by a
large flake of hexagonal boron nitride (hBN). Cross-sectional AFM inspection verifies the
thickness of each exfoliated flake of MoTe$_2$ and graphene used in the devices.

\begin{figure*}[t!]
\begin{center}
\includegraphics[width=\textwidth]{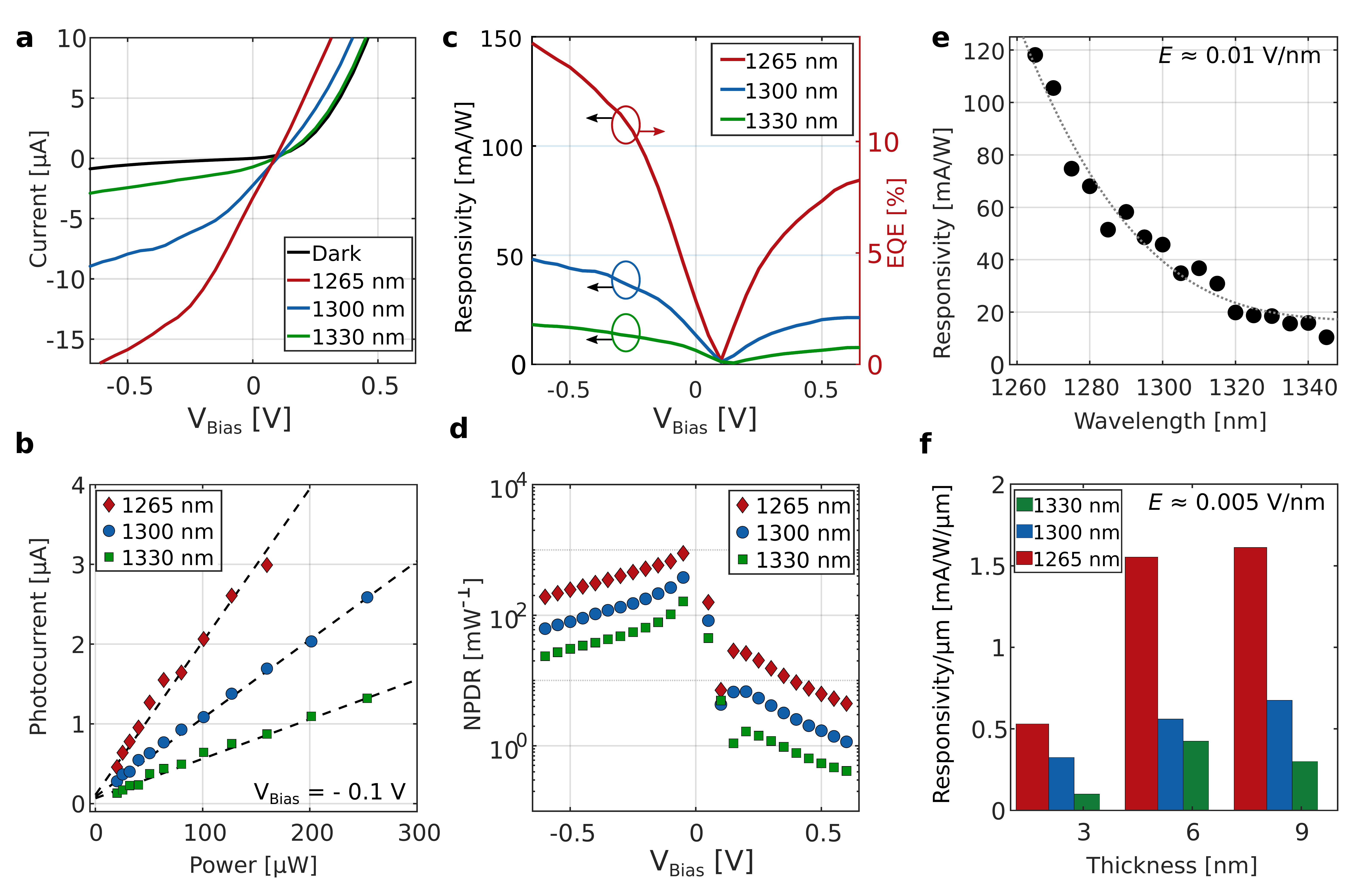}
\caption{\textbf{Steady-state photoresponse of a waveguide photodetector featuring a MoTe$_2$ thickness of 45\,nm.} \textbf{a}, Current-voltage (I-V) curves with and without light for three different
wavelengths (black, dark current; red, 1265\,nm; blue, 1300\,nm; green, 1330\,nm). The optical power is 150 $\mu$W for all three wavelengths. \textbf{b}, Measured photocurrent as a function of the optical
power at $-$0.1\,V bias. \textbf{c}, Measured responsivity (left vertical axis in black)
as a function of applied bias voltage for three different wavelengths. Derived external quantum efficiency (EQE, right vertical axis in red) for input light at 1265\,nm.
A responsivity of 150\,mA/W and a EQE of 14\% are obtained at a low bias voltage
of $-$0.6\,V and for input light at 1265\,nm. \textbf{d}, Normalized photocurrent-to-dark-current ratio
(NPDR) as a function of applied bias voltage. The highest NPDR is calculated to be 1000 mW$^{-1}$ for input light at 1265\,nm and small bias conditions. \textbf{e}, External
responsivity as a function of wavelength for $-$0.5\,V bias. The wavelength
dependent photoresponse agrees with the absorption spectrum of MoTe$_2$. Dots are data and the dashed line is a guide to the eye. \textbf{f}, Comparison of the
responsivity normalized by the device length for devices with different flake
thicknesses and identical electric fields $E\sim 0.005$\,V/nm. Red, blue, and green bars are for light at 1265\,nm, 1300\,nm and 1330\,nm, respectively.
\label{fig3}}
\end{center}
\end{figure*}

\textbf{Electrical Characteristics.}
The electrical behavior of the fabricated devices was
characterized using a pico-ampere precision source to apply biases and read out the current. The metallic contact pads connected to the bottom graphene electrodes were biased
either positively or negatively, while the top contact was
grounded. Figure~\ref{fig2}a illustrates the band diagram of the studied heterostructure under positive and negative bias conditions, respectively. Both MoTe$_2$ and graphene are known to be
lightly p-doped~\cite{Lee2008a}. The Fermi level of Au is aligned and pinned close to the valence band of
MoTe$_2$~\cite{Shin2017a,Nakaharai2016,Wee2017a}. This pinning tends to remain unchanged regardless of the thicknesses of TDMC flakes~\cite{Shin2017a}. Graphene on the other hand forms a
smaller and tunable Schottky barrier after contacting MoTe$_2$~\cite{Wang2016h,Yu2017}. The difference in work
function of Au and graphene leads to a built-in potential in the device. Applying a bias voltage
increases the potential drop across the device that drives the photo-excited carriers. Figure~\ref{fig2}b
shows the current density against the applied bias voltage for devices comprising flakes of
three different MoTe$_2$ thicknesses and lengths, more exactly, 11\,nm and 20\,$\upmu$m, 35\,nm
and 33\,$\upmu$m, and 45\,nm and 40\,$\upmu$m. The measured current is normalized to the effective detector length of each device to account for the size-variations of the used flakes. As expected, the highest current density can be observed for the thinnest device, as a result of the
short resistive MoTe$_2$ channel. Moreover, asymmetric current densities under positive and negative bias conditions, induced by the asymmetric contacting scheme, are visible and more pronounced for thicker devices, most likely originating from a MoTe$_2$ thickness-dependent
Schottky barrier height between graphene and MoTe$_2$, as previously studied and reported in
graphene-TMDC heterostructures~\cite{Yu2013e,Wang2016h}. This asymmetric contacting scheme also manifests
itself in a pronounced photoresponse at zero bias (V$_{Bias}$ = 0 V) resulting from an intrinsic built-in field. Figure~\ref{fig2}c shows current-voltage (I-V) curves of the 45\,nm thick device with and without 1300\,nm light coupled into the waveguide. Without any applied voltage
the device is already capable to efficiently separate photo-excited electron-hole pairs and
to generate a considerable photocurrent of 2 $\upmu$A for 150 $\upmu$W input power, with negligible dark current.

\textbf{Steady-state photoresponse.}
The steady-state photoresponse of the fabricated devices was evaluated using linearly TE-polarized laser light with center wavelength of 1300\,nm coupled into the integrated waveguides via GCs. The coupling loss of a GC was verified to be around 7.5 dB with the help of nearby reference structures featuring identical waveguide and GC designs fabricated on the same chip, but without
detector structures (see Supplementary Section S2 for details). Figure~\ref{fig3}a shows I-V
measurements with and without light coupled into a device with a 45\,nm thick MoTe$_2$ flake. Unless otherwise specified, all
data presented in Fig.~\ref{fig3} are based on this specific device. A pronounced increase in current,
especially under negative bias voltages, is measured when light is coupled in. The power dependence of this photocurrent is shown in Fig.~\ref{fig3}b
for different wavelengths. A linear dependence is observed within the measured
power range. The photoresponsivity is extracted as the ratio of the photocurrent and the incident optical power delivered to the photodetector. It increases
with the applied bias voltage that assists the extraction of carriers. Increasing the applied bias
to moderate values of up to $-$0.6 V, as shown in Fig.~\ref{fig3}c, yields photoresponsivities of 150\,mA/W, 50\,mA/W and 20\,mA/W for wavelengths of 1265\,nm, 1300\,nm and 1330\,nm, respectively. This corresponds to an external
quantum efficiency (EQE) $\eta_{EQE}$ of 14\% at 1265\,nm ($\eta_{EQE} = R \cdot \hbar\omega/q$, $R$ denoting the responsivity, $\hbar$ the reduced Planck constant, $\omega$ the light angular frequency, and $q$ the elementary charge). We intentionally keep the bias voltage low in steady-state
photodetection measurements in order not to damage the devices. In fact, the photocurrent and the EQE are expected to further increase for larger applied bias voltages, limited only by break
down and saturation of absorption (see Supplementary
Section S3 for details on the limits of the responsivity).

Normalized photo-dark-current ratio (NPDR)~\cite{Goykhman2016} is another important performance indicator
of a photodetector. As shown in Fig.~\ref{fig3}d, the NPDR of our devices is in the order of 100 mW$^{-1}$ under negative bias conditions and
approaches 1000 mW$^{-1}$ for shorter wavelengths and small bias
voltages. This performance outperforms graphene-based
photodetectors~\cite{Gan2013c,Mueller2010a} by order(s) of magnitude. We further characterized the dependence of the photoresponse on the wavelength of the incoupled light. As shown in Fig.~\ref{fig3}e, the measured responsivity spectrum agrees with the wavelength dependent absorption of the few-layer MoTe$_2$ flake. It exhibits stronger absorption and hence higher photoresponse for shorter wavelengths.

In Fig.~\ref{fig3}f we compare devices with different MoTe$_2$ thicknesses. The figure shows the responsivities, normalized to device length, for light with wavelengths of 1265\,nm, 1300\,nm and 1330\,nm. The observed trend clearly shows that devices consisting of thicker MoTe$_2$ flakes possess a higher photoresponse, which is due to the higher absorption in the
thicker semiconducting channel.
The behaviour for 1265\,nm (red bars) is the same as for
1300\,nm (blue bars) and 1330\,nm (green bars), despite the overall lower responsivity. These characteristics verify that
the observed photoresponse stems from photo-excited carriers generated by light absorption in the semiconducting MoTe$_2$. Although the bottom graphene contact may also contribute to the
photocurrent via photo-thermionic effects, the observed linear power dependence as well as the wavelength dependent photocurrent in our devices is in stark contrast to the characteristic features of the photo-thermionic effect~\cite{Massicotte2016b} (i.e. superlinear power behavior and wavelength independent photocurrent). We therefore conclude that in the
investigated wavelength range the photo-thermionic contribution of graphene is not significant.

\begin{figure*}[t!]
\begin{center}
\includegraphics[width=\textwidth]{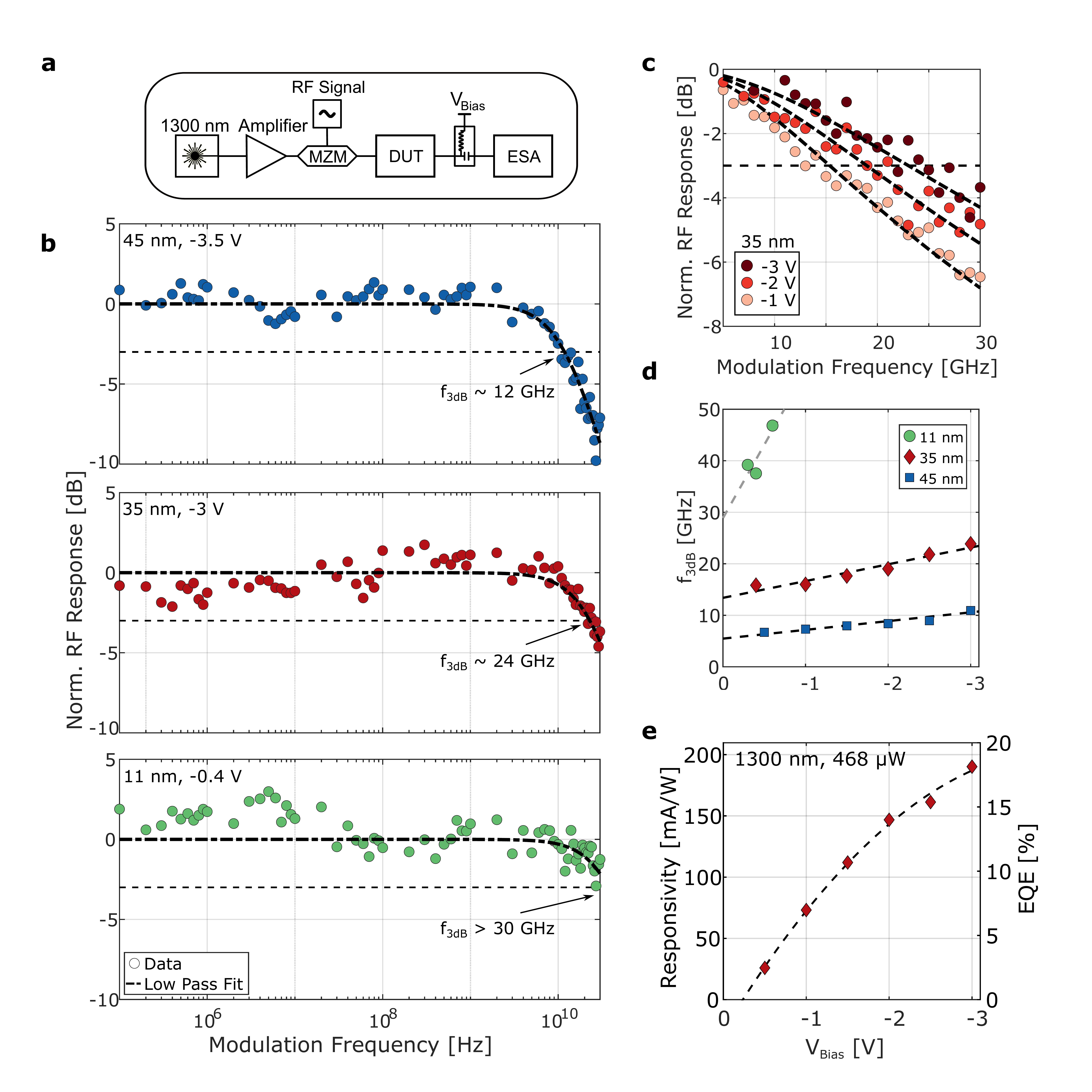}
\caption{\textbf{Dynamic characterization of vertical photodetectors.} \textbf{a}, Frequency
response measurement setup. MZM, Mach-Zehnder modulator; DUT, device under test;
ESA, electrical spectrum analyzer. \textbf{b}, Normalized radio frequency (RF) signal response as
a function of the modulation frequency of the input signal for three devices. A single pole low-pass
filter is used to fit the data points and to extract the 3 dB roll-off frequency $f_{3dB}$ for each device. The thicknesses of the MoTe$_2$ flakes and the applied bias voltages
are indicated in the figures. \textbf{c}, Frequency response of the 35nm thick MoTe$_2$ device for different
bias voltages. \textbf{d}, Dependence of the roll-off frequency $f_{3dB}$ on applied bias voltage and for different MoTe$_2$ flake thicknesses (blue scatters, 45\,nm; red rhombuses, 35\,nm; green dots, 11\,nm). The linear trend indicates that the bandwidth is transit time limited. \textbf{e}, Simultaneously
measured responsivity and corresponding external quantum efficiency as a function of
the applied bias voltage for input light at 1300\,nm with power intensity of 468 $\mu$W.
\label{fig4}}
\end{center}
\end{figure*}

\textbf{High-frequency photoresponse.}
To characterize the speed performance of the devices we used the experimental setup illustrated in
Fig.~\ref{fig4}a. An optical intensity Mach-Zehnder modulator (MZM, u$^2$t MZMO2120) with 30\,GHz electro-optic bandwidth was driven by a radio frequency (RF)
signal from an electrical synthesizer and used to modulate an amplified continuous-wave
laser tunable around 1300\,nm. The modulated laser light was coupled into the device via GCs. A bias-tee was used to apply a direct current (DC) bias to the devices.
The generated RF electrical signals were extracted from the devices with a high-speed
microwave probe and measured with an electrical spectrum analyzer (ESA). The frequency
responses ranging from 100 kHz to 30\,GHz were measured under various bias voltages. The
whole measurement setup was calibrated using a commercially available high-speed
photodetector with a bandwidth of 72\,GHz (u$^2$t XPDV3120R). Figure~\ref{fig4}b shows the measured
frequency response of three photodetector devices with different MoTe$_2$ flake thicknesses. In all three cases the response stays flat from 100 kHz to GHz frequencies and then drop off. A standard low pass filter model was
used to fit the data, revealing the 3 dB roll-off frequency of each measurement. It is evident from Fig.~\ref{fig4}b that devices with thinner MoTe$_2$ exhibit a faster
photoresponse. For the 45\,nm and 35\,nm thick devices we measure a roll-off frequency of 12\,GHz at $-$3.5\,V and 24\,GHz at $-$3\,V bias, respectively. Whereas the bandwidth of the 11\,nm thick MoTe$_2$ device exceeds 30\,GHz already at a low bias of  $-$0.4 V, which is beyond the bandwidth
of the instruments used in the experiment. The extrapolated 3 dB roll-off frequency is nearly 50\,GHz. To the best
of our knowledge, this is the highest reported bandwidth of a TMDC-based photodetector,
outperforming those of previous studies by more than one order of magnitude~\cite{Bie2017,Ma2018a}. For the 35\,nm thick MoTe$_2$ device we show in Fig.~\ref{fig4}c the influence of the bias voltage on the frequency response in the roll-off regime. The 3
dB roll-off frequency increases with the applied bias since photo-excited carriers are separated faster by large electric fields. Figure~\ref{fig4}d plots the roll-off frequencies as a function of the bias voltage for all three devices. It reveals that the bandwidth increases monotonously with bias voltage. This is because the velocity of the carriers increases linearly with applied bias. Furthermore, a high
bias not only increases the bandwidth but also leads to an enhanced photoresponsivity, as a
result of the reduced carrier recombination. As shown in Fig.~\ref{fig4}e, the photoresponsivity increases with the bias
voltage, approaching 200\,mA/W at a bias of $-$3 volt for incident light at 1300\,nm for the 35\,nm thick MoTe$_2$ device, corresponding to an EQE close to 20\%.

It is meaningful to determine the circuital
characteristics of the devices, as the RC products could also impose limits on the time
response. To this end, we experimentally characterized the alternating current (AC) capacitances of the
studied devices (see Supplementary Section S4). Thanks to the compact size of the
devices, the measured capacitances are significantly smaller than those of standard top illuminated
devices~\cite{Massicotte2015c} and are in the range of a few tenths of femtofarads. Thus, the RC limited bandwidths are larger than 100\,GHz if the devices are connected to a 50\,Ohm load. We therefore conclude that the bandwidths of our graphene-MoTe$_2$ photodetectors are mainly limited by the carrier dynamics  (see Supplementary Section S5).

\begin{figure}[]
\begin{center}
\includegraphics[width=23em]{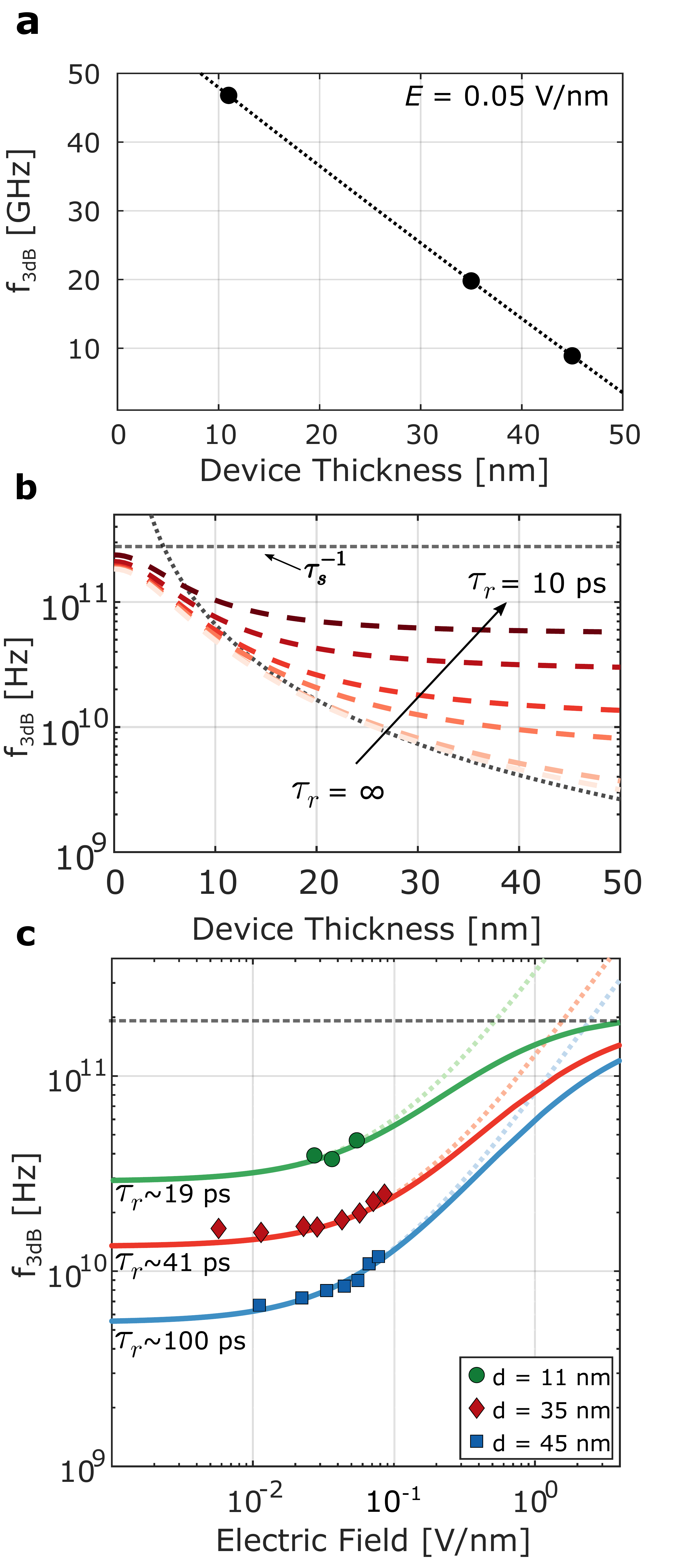}
\caption{\textbf{Comparison of devices with different MoTe$_2$ thicknesses and under different
bias conditions.} \textbf{a}, Roll-off frequencies $f_{3dB}$ for different MoTe$_2$ thicknesses $d$
but identical electric field $E$. \textbf{b}, Predicted roll-off frequencies $f_{3dB}$ for different MoTe$_2$ thicknesses. The dotted line is a plot based on a simple transit time
model. The colorful dashed lines are plots based on the adapted transit time model taking
into account the carrier recombination channel for devices under different bias conditions. \textbf{c},
Measured roll-off frequencies $f_{3dB}$ versus applied electrical field (V$_{Bias}/d$) for
different MoTe$_2$ thicknesses. The indicated fitting parameters $\tau_r$ correspond to the
recombination lifetimes, which are MoTe$_2$ thickness dependent.
\label{fig5}}
\end{center}
\end{figure}

\textbf{Discussion.}
As discussed in Fig.~\ref{fig4}d, the roll-off frequencies depend on the bias voltage. Moreover, for a constant electric field, thin devices show higher roll-off frequencies, as shown in Fig.~\ref{fig5}a. These characteristics disclose drift-diffusive transport to govern the
charge carrier dynamics in our devices. On average, the photo-excited carriers need to drift
over a length $d/2$, where $d$ is the vertical distance between electrodes, i.e. the flake thickness
of MoTe$_2$. The velocity of carriers is given by $\nu d = \mu E \approx \mu V/d$, where $\mu$ is the out-of-plane mobility of carriers, $E$ the electric field across the device and $V$ the applied bias
voltage. The transit time for carriers to be collected can be derived as $\tau_{tr} = d^2/2\mu V$ and the corresponding transit time limited frequency, $f_{3dB}$, is shown as the black dotted line in Fig.~\ref{fig5}b.
This simple model can be used to fit the slope of the linear behavior found in Fig.~\ref{fig4}d, providing an estimate for the out-of-plane mobility of 0.03 cm$^{2}$V$^{-1}s^{-1}$ for all three devices, in good agreement with values reported in the literature~\cite{Yu2013e,Massicotte2015c}. In order to explain the different intercepts of the linear fits at zero bias in Fig.~\ref{fig4}d, however, an additional channel needs to be
considered. As investigated recently by Massicote et al. who employed time-resolved pump-probe measurements to study the carrier dynamics of WSe$_2$ heterostructures, there exists a
loss mechanism stemming from carrier recombination~\cite{Massicotte2015c}. This gives rise to a lower bound of the bandwidth at small bias voltages and the total extraction time $\tau$ can be described by $\tau^{-1} = \tau_{tr}^{-1} + \tau_{r}^{-1}$,
where $\tau_{r}$ denotes the carrier recombination lifetime.

Figure~\ref{fig5}b illustrates the impact this additional parallel channel has on the carrier dynamics. Treating the recombination lifetime $\tau_{r}$ as a variable, the thickness-dependent roll-off frequencies
for various lifetime values are plotted. While in a purely transit time limited case the theoretical maximal
roll-off frequency falls off quickly with increasing thickness $d$ (black dotted line), taking the lifetime $\tau_{r}$ into account significantly
flattens out this drop-off (colored dashed lines). This behavior is more remarkable if the carrier recombination lifetime $\tau_{r}$ is
small, thereby enabling higher roll-of frequencies. In contrast, a slow recombination (large $\tau_{r}$) slows down the overall carrier dynamics.

For completeness, an additional timescale has to be taken into account when looking at the
upper bound of the theoretical bandwidth, namely the lifetime of interfacial processes describing the actual transfer of carriers from the semiconducting channel to the electrodes,
denoted $\tau_{s}$ in Fig.~\ref{fig5}b. While the recombination lifetime $\tau_{r}$ acts as a parallel loss channel, $\tau_{s}$ is a process in series with the photo-carriers extraction. Therefore, it starts to play a role when the transit time reaches a few picoseconds for very thin devices or for devices under very strong electric fields. The lifetime of such a process was reported to be thickness-independent and on the order of 2-5 ps~\cite{He2014a}. This is beyond our experimental study, but shows that the roll-off frequencies may converge to an upper limit beyond 200\,GHz. This predicted ultimate
performance is comparable to the estimated intrinsic bandwidth of graphene, which is about
260\,GHz~\cite{Urich2011}.
We can now compare the extracted roll-off frequencies of different devices as a function of electric field $E$ with the modeled roll-off frequencies using $f_{3dB} = 0.55/\tau$. The  complete carrier rate equation is given by:
\begin{equation*}
    \frac{1}{\tau} = \frac{1}{\tau_s + \frac{d}{2 \mu E}} + \frac{1}{\tau_r}.
\end{equation*}
We find a good agreement between our experimental results and the fit. As expected, thinner channels result in higher roll-off
frequencies when applying the same field across the device. A stronger field raises the roll-off frequency resulting from a shorter transit time of the accelerated photo-excited carriers.
This dependence is more pronounced at high fields and less efficient at low fields, for which the roll-off frequencies start to plateau. Using $\tau_{r}$ as a fitting parameter we obtain a clear dependence on the thickness of the MoTe$_2$ flakes. We find the recombination lifetimes to be
relatively small ($\tau_{r}$ = 19 ps) for thin devices ($d$ = 11\,nm) and large ($\tau_{r}$ = 100 ps)
for thick devices ($d$ = 45\,nm). This trend has been observed previously in transient absorption studies of other
TMDCs~\cite{Cui2014,Shi2013}. The observation that $\tau_{r}$ scales with the thickness $d$ can be attributed to surface defects. Because thin TMDC flakes have a high surface-to-volume
ratio, they are more susceptible to surface defects. Thus, thin flakes are more favorable for fast carrier dynamics, firstly due to their short carrier transit channel and secondly due to their
intrinsically short recombination lifetimes. On the other hand, when the TMDC flake gets even thinner (e.g., $d <$ 10\,nm), the reciprocally increased capacitance may eventually limit the device bandwidth. Our analysis reveals that an optimum thickness around 10\,nm offers the highest bandwidth for a resistively loaded device (see Supplementary Section S5).

Furthermore, since the extraction rate $\Gamma = 1/\tau_{tr}$ increases with thickness, the internal quantum efficiency (IQE), given as the ratio of extracted carriers and total photo-generated carriers, is expected to increase for thin devices. However, as $\tau_{r}$ is small for thin devices, there is also a fast recombination channel that counteracts the carrier extraction. As a result, we find the IQE to be nearly equal (up to $\sim$\,40\% for our measured voltage range) for the different thicknesses, and only dependent on the applied field (more details in Supplementary Section S3). The reason why a thin device typically exhibits a lower responsivity or a lower EQE than those of thicker devices, can be attributed to the reduced photoactive material. Hence, it appears that a low responsivity is the price to pay for a high
bandwidth. However, the responsivity of our waveguide-integrated devices can readily be improved by simply increasing the length of the flake on the waveguide. Thus, our proposed design concept shows a viable scheme to overcome the common trade-off between high (external) efficiency and fast intrinsic photoresponse.

In conclusion, our findings and results have profound impact on the understanding and development of practical TMDC optoelectronic devices. The presented waveguide coupled
vertical heterostructure device concept paves a way to boost the speed performance of
TMDC-based photodetectors to the same order of magnitude as those of e.g. graphene,
which, although highly promising for very high bandwidth applications, suffers from
comparably weak light-matter interaction and large dark currents. Both of these issues can be
addressed by integrating TMDCs and their heterostructures with
silicon integrated photonics. The presented waveguide-integrated device design extends the
potential of TMDCs for practical optoelectronic devices particularly in the fields of
high-speed applications such as high-data-rate optical interconnects operated at standard
telecom wavelengths on silicon photonics platforms.\\

\section*{Methods}
Device fabrication. Photodetectors were fabricated on a standard silicon-on-insulator (SOI)
wafer. Buried silicon waveguides with dimensions of an effective width $w$ = 400\,nm and a
height $h$ = 220\,nm were first built by using the LOCal Oxidation of Silicon (LOCOS)
technique (see Supplementary Information, S2). Grating couplers (GCs) were produced by a
shallow etching of silicon. A top 5\,nm thick SiN dielectric layer was then deposited by
atomic layer deposition for an electrical isolation from the silicon layer underneath. Next to
the waveguide, bottom metallic pads which were used to contact the graphene electrode were 
subsequently defined by electron-beam lithography, evaporation of 5\,nm Ti and 50\,nm Au,
and a lift-off process. Mechanical exfoliation was employed to obtain crystalline flakes of
MoTe$_2$, graphene and hBN, which were identified with an optical microscope and whose
thicknesses were characterized with an atomic force microscope (AFM). The graphene MoTe$_2$ heterostructure were stacked by using a polymer-based pick-up technique with a
polydimethylsiloxane (PDMS) polypropylene carbonate (PPC) stamp, transferred to the
device chips, and aligned to the silicon waveguides with the help of the micromechanical
stage of a SUSS MJB4 mask aligner. 200\,nm wide and 20\,nm thick top Au contact pads were formed again by
electron-beam lithography, metal evaporation, and a lift-off process. The whole devices were
finally encapsulated by hBN flakes. The measurements were performed at ambient
conditions at room temperature.

\begin{acknowledgements}
This research was supported by the Swiss National Science Foundation (grant no. 200021\_165841). K.W. and T.T. acknowledge support from the Elemental Strategy Initiative
conducted by the MEXT, Japan, A3 Foresight by JSPS and the CREST
(JPMJCR15F3), JST.
This work was carried out partially at the Binnig and Rohrer Nanotechnology Center and the FIRST Center for Micro- and Nanotechnology at ETH Zurich.
\end{acknowledgements}

\section*{Author contributions}
N.F. and P.M. conceived the concept, designed and fabricated the devices, designed and performed the experiments, and analyzed the data. Y.S. contributed to the experiments. A.E. contributed to the device fabrication. T.T. and K.W. synthesized the hBN crystals. N.F., P.M., J.L., and L.N. co-wrote the manuscript, with support from all authors.\\
N.F. and P.M. contributed equally.

\bibliographystyle{naturemag}


\end{document}